\documentclass[reprint,
amsmath,amssymb]{elsarticle}

\usepackage{graphicx}
\usepackage{subfig}
\graphicspath{{./figures/}}
\usepackage{dcolumn}
\usepackage{bm}
\usepackage[mathlines]{lineno}
\usepackage[export]{adjustbox}
\usepackage{color}
\usepackage{tabularx,array,booktabs}
\usepackage{caption}
\usepackage{mathtools}
\usepackage{soul}

\newcommand{\X}{\boldsymbol{X}}
\newcommand{\x}{\boldsymbol{x}}
\newcommand{\up}{\boldsymbol{u}}
\newcommand{\U}{\boldsymbol{U}}

\newcommand{\N}{\nabla_{\boldsymbol{X}}}
\newcommand{\n}{\nabla}

\newcommand{\F}{\boldsymbol{F}}
\newcommand{\V}{\boldsymbol{V}}
\newcommand{\R}{\boldsymbol{R}}

\newcommand{\Hp}{\boldsymbol{\Psi}}
\newcommand{\Psirvc}{\Phi}
\newcommand{\Phirvc}{\Psi}
\newcommand{\rvc}[1]{\textcolor{black}{#1}}

\begin{document}

\begin{frontmatter}

\title{
 Elastic Wave Near-Cloaking}

\author[polimi]{Davide Enrico Quadrelli\fnref{myfootnote}}
\fntext[myfootnote]{davideenrico.quadrelli@polimi.it}
\author[imperial,imperial2,imperial3]{Richard Craster}
\author[femto]{Muamer Kadic}
\author[polimi]{Francesco Braghin}

\address[polimi]{Department of Mechanical Engineering, Politecnico di Milano, via La Masa 1, 20156 Milano, Italy}
\address[imperial]{Department of Mathematics, Imperial College London, London SW7 2AZ, UK}
\address[imperial2]{Department of Mechanical Engineering, Imperial College London, London SW7 2AZ, UK}
\address[imperial3]{UMI 2004 Abraham de Moivre-CNRS, Imperial College London, London SW7 2AZ, UK}
\address[femto]{Institut FEMTO-ST, UMR 6174, CNRS, Universit\'{e} de Bourgogne Franche-Comt\'{e}, 25000 Besan\c{c}on, France}
\date{\today}

\begin{abstract}
 Cloaking elastic waves has,
 in contrast to the cloaking of electromagnetic waves, remained a fundamental challenge: the 
 latter successfully uses the invariance of Maxwell's equations, from which the field of transformational optics has emerged, whereas the elastic Navier equations are not invariant under coordinate transformations.
  Our aim is to overcome this challenge, at least in practical terms, and thereby unlock applications in mechanics, ultrasound, vibration mitigation, non-destructive evaluation and elastic wave control. We achieve near-cloaking by recognising that, despite the lack of invariance,  
  a decoupling into a system of form invariant potential equations together with a quantifiable approximation, can be used effectively in many cases to control the flow of elastodynamic waves. Here, in particular we \rvc{focus} on the efficiency and practicability of the proposed near-cloaking which is illustrated using carpet cloaks to hide surface defects from incoming 
  compressional and shear in-plane waves and from surface elastic Rayleigh waves. 
 

\end{abstract}

\begin{keyword}
Metamaterials, 
\end{keyword}

\end{frontmatter}
\section{Introduction}

Motivated by the successes of cloaking for electromagnetic waves there is considerable interest in mirroring this success for elastic waves. Unfortunately, elastic waves for the full vectorial system have several complications, and differences, from their electromagnetic counterparts, for instance the shear and compressional wavespeeds being different
 and the inherent coupling of wave-types upon reflection from most interfaces amongst other reasons \cite{kadic2020}: achieving practical cloaking remains an open challenge in the literature.  The important technical step underlying the  electromagnetic transformations is the invariance of the Maxwell system, but 
 their counterparts, the elastic Navier equations, are generally not form invariant under general coordinate transformations \cite{Milton2006,Norris2008,Norris2011,diatta2016scattering,buckmann2014elasto}.
  The critical issue is the choice of gauge for the transformation \cite{kadic2020} and, depending upon the precise transformation, it always produces restricted and impractical materials. 
  
  Several ideas have been advanced to sidestep this restriction and allow for cloaking, for instance using a pentamode assumption which in principle decouples the longitudinal and transverse waves by canceling the latter \cite{Milton1995,Kadic2012}. An alternative approach is to introduce Cosserat cloaks that have non-symmetric effective elastic tensors \cite{Norris2011}, or in a similar vein Willis media \cite{Milton2006}. In addition Parnell et al \cite{parnel2012,parnell2012employing,parnell2012nonlinear} have proposed to use nonlinear cloaks based on hyperelastic neo-Hookean materials where the induced deformation in the geometry would actually come from a real inflation and where the material would by itself create the required parameters for the cloak. 
  In many of these examples, experiments are limited by the necessity of having inhomogeneous and anisotropic distribution of parameters of the cloak; the only truly Cosserat-like cloak was only proposed  recently \cite{Xu2020}. It is possible to partially sidestep the issue if the elastic body is a thin-elastic plate for which flexural waves dominate; there is a large literature on this subcase where the scalar Kirchhoff-Love plate equations are used, \cite{Misseroni2019} but even in this limited case cloaking is still an open issue. For the vectorial elastic case we consider here even approximate and narrow-band cloaking has not been achieved and experiments are limited to the quasi-static case \cite{buckmann2014elasto}.
  
 The literature on transformational optics, and its generalisations, is now vast and it is a widespread tool in both optics and acoustics; the concepts can be traced back to calculations in  hydrodynamics \cite{Lamb1895}, and \cite{Dolin1961} for coating layers to prevent distortion. In numerical simulation, transformations \cite{Imhoffl1990,Nicolet1994} created effective infinite domains that led to the PMLs (perfectly matched layers) \cite{Berenger1994} now widely used.
 The work of Pendry \cite{Pendry2006} and Leonhard \cite{Leonhardt2006} in designing invisibility cloaks that act to hide an object from outside radiation by producing zero scattering, and zero interaction of the wave with the object, led to an explosion of activity with the idea of invisibility or cloaking then adapted to many physical, scalar problems such as electromagnetic/optical invisibility \cite{Pendry2006,Leonhardt2006}; cloaks for water waves \cite{Farhat2008,Farhat2009}; airborne sound \cite{Cummer2008,cummer2016,Popa2011,Torrent2006,Torrent2006b,Torrent2008a,Torrent2008b,Norris2008};  vibration cloaks \cite{Milton2006,Farhat2008,Farhat2009,Farhat2009b,Stenger2012,buckmann2014elasto,kadic2020,Misseroni2019}; and  thermal cloaks \cite{Guenneau2012,Schittny2013,Ji2019}. In parallel with the idea of infinite transformations such as conformal cloaking \cite{Leonhardt2006} there were strides to make the transformations more realistic using the  truncation of the infinite transformation to a  finite one \cite{LiPendry2008}. An alternative approach called scattering cancellation, or neutral inclusions, has also been used for long wavelength considerations \cite{alu2005achieving,Schittny2013,buckmann2014elasto,Ji2019}. 
  Carpet cloaks, based around the idea of quasi-conformal mappings \cite{LiPendry2008} have been particularly popular as such cloaks allow one to effectively hide defects or imperfections in a surface, and have implementations in optics \cite{Zhang2011b,Chen2011}. Overcoming the current limitations in full elasticity, even if not for perfect cloaking, will open up many of these developments in optics to their counterparts in elasticity.

There are also numerous fabrication issues with making cloaks, and those for the simple elastic plates often involve multiple layers and media \cite{Stenger2012} to mimic anisotropy and on a practical level we observe that making isotropic metamaterials is easier than aiming for designed anisotropy, and that a continuously changing material (adiabatic) is more practicable than any laminating technique or any multiscale approach. The possibility of using isotropic materials to control the propagation of elastic waves was previously mentioned in \cite{chang2011controlling}, were general transformations are idealized by a series of local affine ones point-by-point, and in \cite{gao2018manipulate}, where the material specification in the cloak is obtained by enforcing the invariance of potential energy after the transformation. 

Following a different path, in section \ref{sec:theory} we cover the transformation required for the medium to reroute wave motion, and then turn to the elastic Navier equations \cite{banerjee2011introduction}, decouple it into potential equations and apply the formalism of the well-known quasi-conformal cloaking. In doing so, we clearly see the terms that are omitted when we make this transformation and we are then able to quantify the magnitude of this omission; hence we aim for near-cloaking, i.e. not perfect cloaking, but \rvc{with the ability} to operate broadband. Results given in section 
\ref{sec:results} show these ideas applied to carpet cloaking of a defect upon an otherwise flat traction-free surface, as shown in Fig. \ref{Figure1}, with in-plane 
 compressional and shear waves incident upon the defect. Reflections are analyzed and considered both as the defect size and frequency vary, in light of a performance metric than is introduced based on the magnitude of the aforementioned omitted terms. Dependence of performance with respect to angle of incidence is also considered. \rvc{An important class of problems involve surface Rayleigh waves that propagate along flat surfaces, these are localised} to the surface with exponential decay in depth, and it is natural to consider whether the carpet cloaks we design \rvc{also} enable them to propagate without loss and we analyse this in section \ref{sec:rayleigh}. Finally, we draw together concluding remarks in section \ref{sec:concluding}.

\begin{figure}[ht!]
\centering
\includegraphics[width=0.65\textwidth]{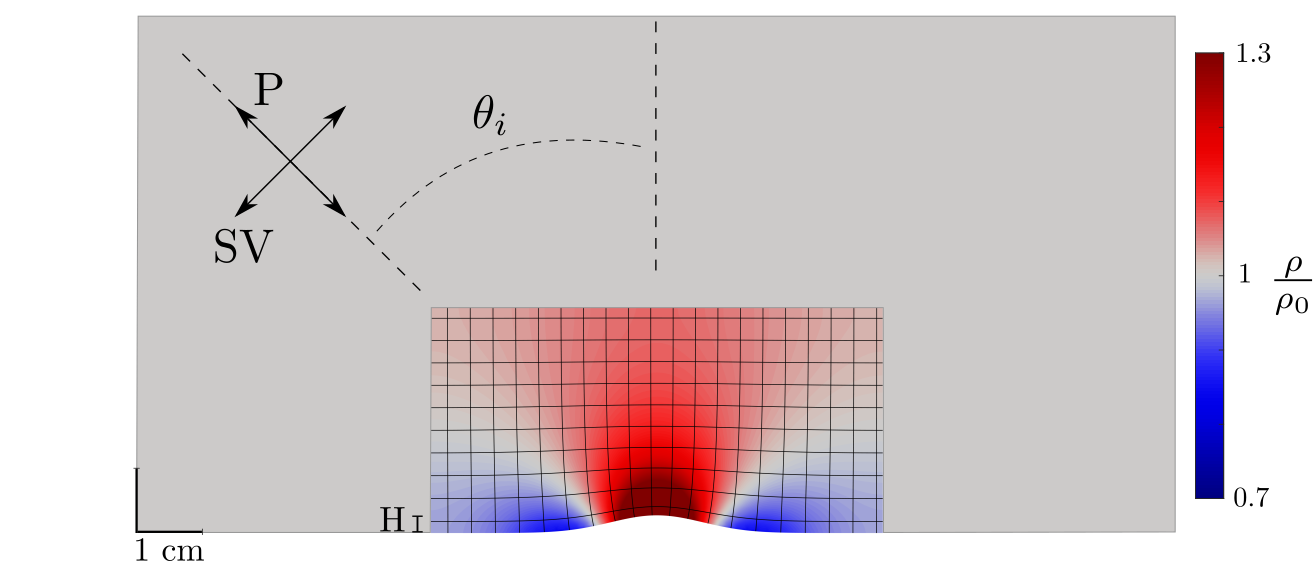}
\caption{Carpet-cloak: A defect on, an otherwise flat, traction-free surface is subjected to incoming compressional $P$ or shear $SV$ waves (with angle of incidence $\theta_i$). Shown is the maximum height of the defect, $H$, the 
calculated metric deformation, and the corresponding local normalised density  $\rho/\rho_0$;  
the color bar indicates values of normalised density.}
\label{Figure1}
\end{figure}


\section{Theory}
\label{sec:theory}

\subsection{Transformation}
\label{sec:transformation}
\noindent We consider a point-wise deformation from the undeformed virtual domain $\X \in \Xi$ to the deformed \rvc{domain} $\x= \boldsymbol{\chi}(\X) \in \xi$ with  the deformation gradient defined by $\F= \nabla_{\boldsymbol{X}}\x$ with $\F^{-1}= \n \X$, where $\N$ represents the differential operator \rvc{for the gradient} in the original undeformed coordinates, whereas $\n$ refers to the same operator in the deformed configuration \cite{Norris2011}. Similarly  capital and plain letters will \rvc{henceforth  distinguish} undeformed and deformed configurations respectively. The Jacobian $J=\det\F$ represents the ratio of volume elements between deformed and undeformed configurations. We note that the polar decomposition of the deformation gradient reads $\F=\V \R$, with $\R$ a local rigid rotation ($\R^T\R=\R\R^T=\boldsymbol{I}, \det \R=1)$ and $\V$ the symmetric positive definite left stretch tensor satisfying $\V^2=\F\F^T$.\\

\noindent A conformal map is defined as a map that preserves angles \cite{Leonhardt2006,Mahan2002} and in the language of finite deformation, it can be generally described as a homogeneous stretch superimposed to a local rigid rotation:
\begin{equation}
	\F=n \boldsymbol{I}\R
\end{equation}
with $\V=n\boldsymbol{I}$ and $n$ being the stretch experienced isotropically by every fiber of space\rvc{; } the change of volume \rvc{ is} $J=n^m$, \rvc{where $m$ is} the dimension of the problem. Under the transformation, the Laplacian in the undeformed configuration transforms  according to \cite{Greenleaf2003} as:
\begin{equation}
\N^2=J \n \cdot (J^{-1}\V^2\n )
\end{equation}
 that, for the conformal map, is
\begin{equation}
	\N^2=n^m \n \cdot (n^{2-m}\n ),
	\label{eq:Three}
\end{equation}
\rvc{In the following sections we apply this transformation to the elasticity equations. }
\subsection{Antiplane Motion (SH waves)}
\rvc{First, we set the scene using anti-plane shear waves} where the displacement $U_3(X_1,X_2)$ is governed by the two-dimensional scalar wave equation:
\begin{equation}
	\bar{\nabla}_{\X} \cdot (\mu_0\bar{\nabla}_{\X} U_3)=\rho_0\frac{\partial^2 U_3}{\partial t^2}
	\label{eq:Four}
\end{equation}
with $\bar{\nabla}_{\X}$ as the differential operator in two dimensions, $\mu_0$ and $\rho_0$ the shear modulus and the mass density, respectively; \rvc{for homogeneous media \eqref{eq:Four} is:}
\begin{equation}
\bar{\nabla}_{\X}^2U_3=\frac{1}{c_{S0}^2}\frac{\partial^2 U_3}{\partial t^2}
\end{equation}
with shear wavespeed $c_{S0}=\sqrt{\displaystyle\frac{\mu_0}{\rho_0}}$.
Applying the conformal map \eqref{eq:Three} then gives
\begin{equation}
\bar{\n}^2 u_3=\frac{1}{n^2c_{S0}^2}\frac{\partial^2 u_3}{\partial t^2}
\end{equation}
 where $u_3(\x(\X))=U_3(\X)$.\rvc{ It is now possible to deduce a new material, occupying the domain $\xi$, with shear modulus $\mu$ and density $\rho$ such that it mimics the virtual domain}. The wave equation governing its anti-plane motion reads:
\begin{equation}
	\bar{\n}\mu\cdot\bar{\n} u_3 + \mu \bar{\n}^2u_3=\rho\frac{\partial^2 u_3}{\partial t^2}
	\label{eq8}
\end{equation}
and if the gradient of $\mu$ is null, then it is sufficient that
$$\displaystyle\frac{\mu}{\rho}=n^2c_{S0}^2$$
for the medium to mimic the virtual domain. In particular, selecting material properties 
$\mu=\mu_0$ and $\rho=
{\rho_0}/{n^2}$ gives the perfect conformal cloak and this process mirrors that in electromagnetism or acoustics \cite{li2008hiding,amirkhizi2010stress}. 

\subsection{In Plane Motion (P and SV waves)}
We now turn our attention to the full
vector equations of elastodynamics that do not allow the neat process in the previous section to be performed exactly. The vector displacement $\U$ in homogeneous media obeys the elastodynamic Navier wave equation  \cite{graff2012wave}:
\begin{equation}
    (\lambda_0+2\mu_0)\N(\N \cdot \U) - \mu_0 \N \times \N \times \U = \rho_0 \frac{\partial^2 \U}{\partial t^2}
\end{equation}
with $\lambda_0$ and $\mu_0$ the Lamè parameters of the material. The Helmholtz decomposition 
\begin{equation}
\U= \N \Psirvc + \N \times \Hp, \quad \N \cdot \Hp=0
\end{equation}
using the  
scalar compressional $\Psirvc$, and vector shear $\Hp$, potentials
 gives two wave equations:
\begin{equation}
 \N^2\Psirvc=\frac{1}{c_{P0}^2}\frac{\partial^2 \Psirvc}{\partial t^2}, \qquad
 \N^2\Phirvc_i=\frac{1}{c_{S0}^2}\frac{\partial^2 \Phirvc_i}{\partial t^2} \qquad i=1,2,3
\end{equation}
with compressional wavespeed $c_{P0}^2=
({\lambda_0+2\mu_0})/{\rho_0}$ and shear wavespeed $c_{S0}^2=
{\mu_0}/{\rho_0}$. At first sight these appear as decoupled wave equations and naively applying the same process, using the conformal map, as in anti-plane elasticity appears attractive. In terms of an exact solution this idea is flawed as traction-free boundary conditions implicitly couple the potentials. 

We consider plane wave propagation in the $X_1, X_2$ plane: an harmonic P-wave and shear waves are obtained from
\begin{equation}
	\Psirvc(\X, t)=\Psirvc_0 e^{j(\omega t - \boldsymbol{\kappa}\cdot \X)}, \qquad 
	\Hp (\X,t)=\Hp_0 e^{j(\omega t - \boldsymbol{\kappa} \cdot \X )}
\end{equation}
respectively.
We align our axes so that for plane wave propagation no gradients are observed in the third direction, and the problem is then essentially two-dimensional, and the wave equation for the scalar potential becomes:
\begin{equation}
    \bar{\nabla}_{\X}^2\Psirvc=\frac{1}{c_{P0}^2}\frac{\partial^2\Psirvc}{\partial t^2}
\end{equation}
Moreover, the displacements associated with shear waves,  are expressed in terms of $\Hp$ as
\begin{equation}
	\U=\N \times \Hp=\begin{bmatrix} \Phirvc_{3,2}\\
	-\Phirvc_{3,1}\\
	\Phirvc_{2,1}-\Phirvc_{1,2}
    \end{bmatrix}	
\end{equation}
For a SV wave there is  no antiplane motion thus:
\begin{equation}
    \Phirvc_{2,1}-\Phirvc_{1,2}=0
\end{equation}
 and without loss of generality $\Phirvc_1$ and $\Phirvc_2$ are taken zero, thus the vector wave equation governing plane shear waves is the two-dimensional scalar wave equation in $\Phirvc_3$:
\begin{equation}
	 \bar{\nabla}_{\X}^2\Phirvc_3=\frac{1}{c_{S0}^2}\frac{\partial^2 \Phirvc_3}{\partial t^2}
\end{equation}
To summarise, the equations for in-plane motion are:
\begin{equation}
\bar{\nabla}_{\X}^2\Psirvc=\frac{1}{c_{P0}^2}\frac{\partial^2\Psirvc}{\partial t^2}, \qquad
\bar{\nabla}_{\X}^2\Phirvc_3=\frac{1}{c_{S0}^2}\frac{\partial^2 \Phirvc_3}{\partial t^2}.
\end{equation}
As previously seen in section \ref{sec:transformation}, after transformation, these equations become 
\begin{equation}
 \bar{\n}^2\phi=\frac{1}{n^2c_{P0}^2}\frac{\partial^2 \psi}{\partial t^2}, \qquad
 \bar{\n}^2 \psi_3=\frac{1}{n^2c_{S0}^2}\frac{\partial^2 \phi_3}{\partial t^2}.
\label{Potentials}
\end{equation} 
To find the corresponding material properties that, in the physical domain, mimic the wave propagation occurring in  virtual domain, we now consider the elastodynamic wave equations for general inhomogeneous media:
\begin{equation}
\begin{split}
    &\n \lambda(\n \cdot \up) + \n \mu \cdot [\n \up + (\n \up)^T]+\\
    &+(\lambda + 2 \mu) \n(\n \cdot \up) - \mu \n \times \n \times \up
    =\rho \frac{\partial^2{\up}}{\partial t^2}.
    \end{split}
    \label{inomog}
\end{equation}
If the gradient of the Lamé parameters is null, upon the substitution:
\begin{equation}
	\up=\n \phi + \n \times \boldsymbol{\psi} 
\end{equation}
the following is obtained:
\begin{equation}
\begin{split}
    &(\lambda+2\mu)\nabla(\nabla^2\phi)-\rho\frac{\partial^2}{\partial t^2}\nabla\phi=\\
   &=-\left[ \mu \nabla \times \nabla^2 \boldsymbol{\psi}-  \rho \frac{\partial^2}{\partial t^2}(\nabla \times \boldsymbol{\psi}) \right]
    \label{eq24}
\end{split}
\end{equation}
This is rearranged as:
\begin{equation}
    \begin{split}
        &\nabla \left[ (\lambda+2\mu) \nabla^2 \phi -\rho \frac{\partial^2 \phi}{\partial t^2} \right] + \nabla \rho \frac{\partial^2 \phi}{\partial t^2}=\\
        &=\nabla \times \left[ \mu \nabla^2 \boldsymbol{\psi} - \rho \frac{\partial^2 \boldsymbol{\psi}}{\partial t^2} \right] + \nabla \rho \times \frac{\partial^2 \boldsymbol{\psi}}{\partial t^2}
    \end{split}
    \label{eq25}
\end{equation}
thus, if the gradient of density can be considered negligible, then the identity is satisfied by
\begin{equation}
    \begin{split}
        &(\lambda+2\mu) \nabla^2 \phi -\rho \frac{\partial^2 \phi}{\partial t^2}=0\\
        &\mu \nabla^2 \boldsymbol{\psi} - \rho \frac{\partial^2 \boldsymbol{\psi}}{\partial t^2}=0
    \end{split}
\end{equation}
and the same decomposition in scalar and vector potentials is seen to be applied as in the case of the virtual domain. The cloak for plane waves can then be performed by scaling both longitudinal and transversal wave speeds according to the local metric change of the transformation.
The displacements will be then associated to the gradient of $\phi$ and to the curl of $\boldsymbol{\psi}$.
\begin{equation}
	\up=\n \phi + \n \times \boldsymbol{\psi} 
\end{equation}
since for the chain rule $\n=\F^{-T}\N$, then, recalling that $\V \in \mathrm{Sym}^+$ is symmetric and $\R$ is orthogonal:
\begin{equation}
	\up=\F^{-T}\U =\V^{-1}\R\U=\frac{1}{n}\R\U
\end{equation}
This means that the displacements assume the same local rotation of the medium and are simply scaled according to the local stretch.\\

Clearly an assumption has been made above regarding the density gradient and we now assess and quantify the impact, and limitations, of doing so, by obtaining a global performance index for the cloak. 
Turning to Eq. (\ref{eq25}): considering the term proportional to $\nabla \rho$ as being negligible, is also interpreted as adding an extra term to each side of (\ref{eq24}), such that the the left hand side is rewritten as:
\begin{equation}
    (\lambda+2\mu)\n (\n^2\phi)-\rho \frac{\partial^2}{\partial t^2}\n \phi -\n \rho \frac{\partial^2\phi}{\partial t^2}
    \label{eq30}
\end{equation}
while the right hand side becomes:
\begin{equation}
    -\left[\mu \nabla \times \nabla^2 \boldsymbol{\psi}- \rho \frac{\partial^2}{\partial t^2}(\nabla \times \boldsymbol{\psi}) -\n \rho \times \frac{\partial^2 \boldsymbol{\psi}}{\partial t^2} \right]
    \label{eq31}
\end{equation}
We use the relation for plane waves in homogeneous media $\phi=\hat{\phi}e^{i (\omega t- \boldsymbol{\kappa}_P \cdot \boldsymbol{x})}$  to make estimates of the modulus of the last two terms in (\ref{eq30}): 
\begin{equation}
    \left|\rho \frac{\partial^2}{\partial t^2}\n \phi \right| \approx \rho_0 \omega^2 \kappa_P \hat{\phi}; \qquad \left|\n \rho \frac{\partial^2\phi}{\partial t^2} \right| \approx<\nabla \rho >\omega^2 \hat{\phi}
\end{equation}
having introduced $<\nabla \rho>$, i.e. the averaged value of the gradient of density over the surface of the cloak, as an estimate of $\nabla \rho$. The order of magnitude of the ratio between the two is:
\begin{equation}
    O\left( \frac{<\nabla \rho >/\rho_0}{\kappa_P} \right)=O\left( \frac{<1/n^2>c_P}{2\pi f}\right)
    \label{magP}
\end{equation}
which shows this added extra term can be reasonably considered negligible in the limit where $\kappa_P$ becomes very large and/or the fluctuation in density with respect to background value (i.e. the fluctuations of metric change) are small. A similar approach applies for the last two terms in (\ref{eq31}): assuming $\boldsymbol{\psi}=\hat{\boldsymbol{\psi}}e^{i (\omega t- \boldsymbol{\kappa}_S \cdot \boldsymbol{x})}$ the modulus of the related terms become:
\begin{equation}
    \left|\rho \frac{\partial^2}{\partial t^2}(\nabla \times \boldsymbol{\psi}) \right| \approx \rho_0 \omega^2 \kappa_S |\hat{\boldsymbol{\psi}}|; \quad \left|\n \rho \times \frac{\partial^2 \boldsymbol{\psi}}{\partial t^2} \right| \approx<\nabla \rho >\omega^2 |\hat{\boldsymbol{\psi}}|
\end{equation}
which means the ratio scales as
\begin{equation}
    O\left( \frac{<\nabla \rho >/\rho_0}{\kappa_S} \right)=O\left( \frac{<1/n^2>c_S}{2\pi f}\right)
    \label{magS}
\end{equation}
As a consequence, for a given transformation the cloak is expected to have increasingly good performances for all frequencies above a certain cutoff, which is set from the conditions expressed by requiring the adimensional ratios (\ref{magP}) and (\ref{magS}) being small. Note that, since the wavelength of P waves is always longer than that of shear waves ($c_P>c_S$), the condition set by $\kappa_P$ is the one that is more demanding. The difference between (\ref{magP}) and (\ref{magS}) only depends on the ratio of wave speeds, that in turn depends only on Poisson's ratio. Reducing $\nu$ results in a reduced difference between $c_P$ and $c_S$, thus reduced difference between the performance of the cloak with respect to the two types of waves.\\

Note that the same distribution of refraction index could be in principle obtained keeping fixed the density and scaling both the Lamè parameters or even changing all the three material parameters, as proposed in previous works \cite{chang2011controlling}, \cite{gao2018manipulate}. However, following the previous analysis, it is clearly less convenient to change both Lamé parameters and density at the same time, because further approximations will be introduced in passing from (\ref{inomog}) to (\ref{eq24}), other than those that were just analyzed. Moreover, if modulation of the stiffness properties is adopted instead of the density, according to (\ref{eq8}), some non idealities will be introduced in the anti-plane solution also, which is instead perfect when density modulation only is considered.

\section{Results and discussion}
\label{sec:results}
\subsection{Carpet cloaking: In-plane wave incidence}
\label{sec:carpet}

\begin{figure*}[ht!]
    \centering
    \includegraphics[width=\textwidth]{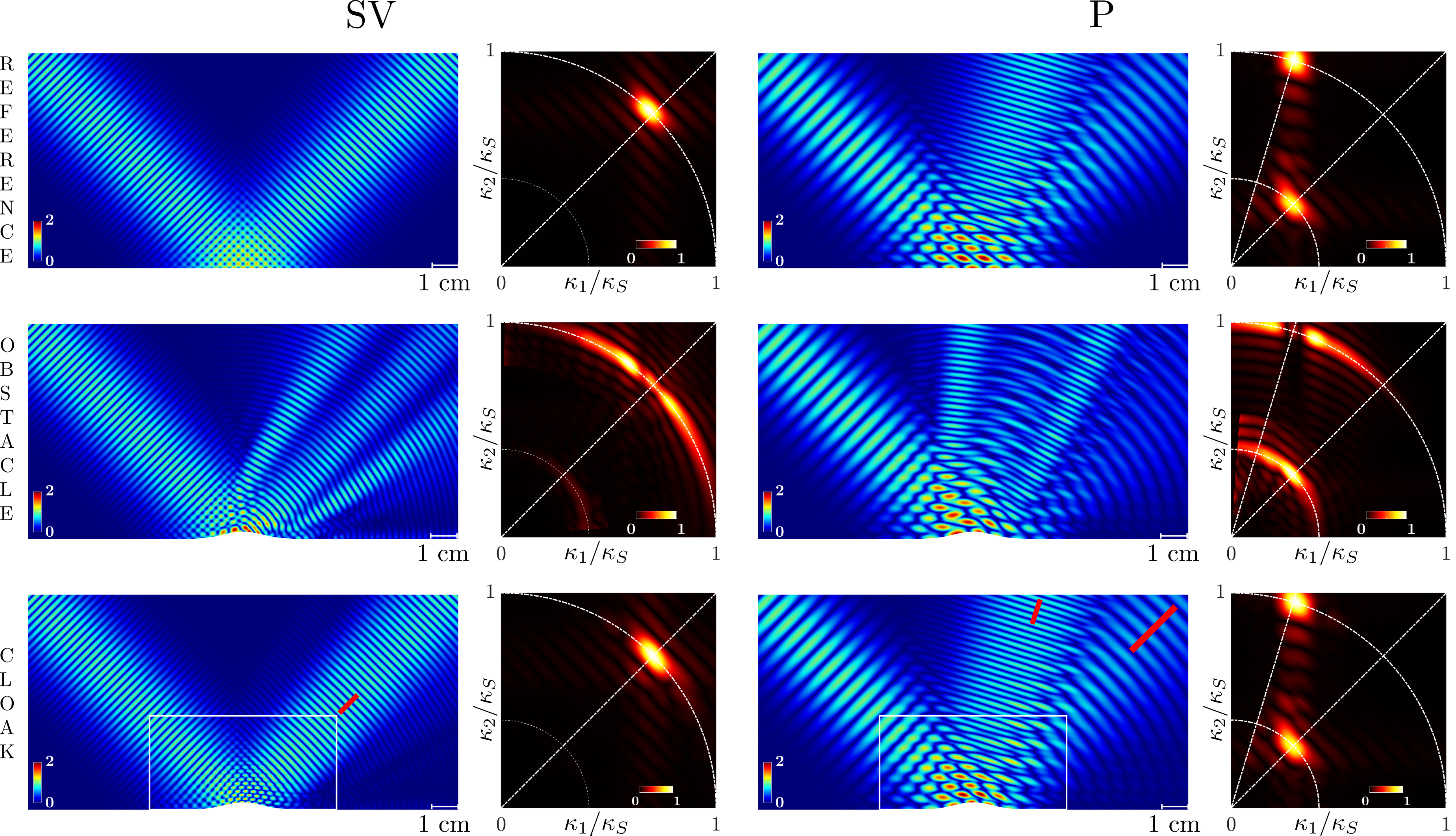}
    \caption{Numerical results for the carpet cloak in terms of field distribution and FFT of scattered fields. The wavenumber axis are normalized with respect to $\kappa_S=\omega/c_S$, such that scattered S waves appear as spots on the unit circle and P waves as spots on the circle with radius $\kappa_P/\kappa_S=c_S/c_P\approx 0.4$. Expected reflection angles are marked with dashed white lines. Three polarizations are shown for 3 configurations (reference: flat mirror, obstacle and cloak). The fields are normalized with respect to the amplitude of the incident wave.}
    \label{Figure2}
\end{figure*}
We now exploit the insight gained from our theory 
 to obtain a carpet cloak working for both in-plane and anti-plane motion. A height defect with Gaussian shape on an otherwise flat free surface is probed by an incident wave. The conformal map linking such deformed domain to the reference flat surface is shown in Figure \ref{Figure1} and used to evaluate the local metric deformation $n$. As discussed earlier such a stretch corresponds to the refraction index of both P and S wave speeds, and the cloak is simply obtained by scaling the density of the material around the obstacle. A nuance is that the conformal map is defined on the whole semi-infinite space, whereas the cloak has to be finite in size. This issue is simply addressed by  truncating the cloak sufficiently far from the obstacle to let the stretch approach unity.
 \begin{figure*}[ht!]
     \centering
     \includegraphics[width=\textwidth]{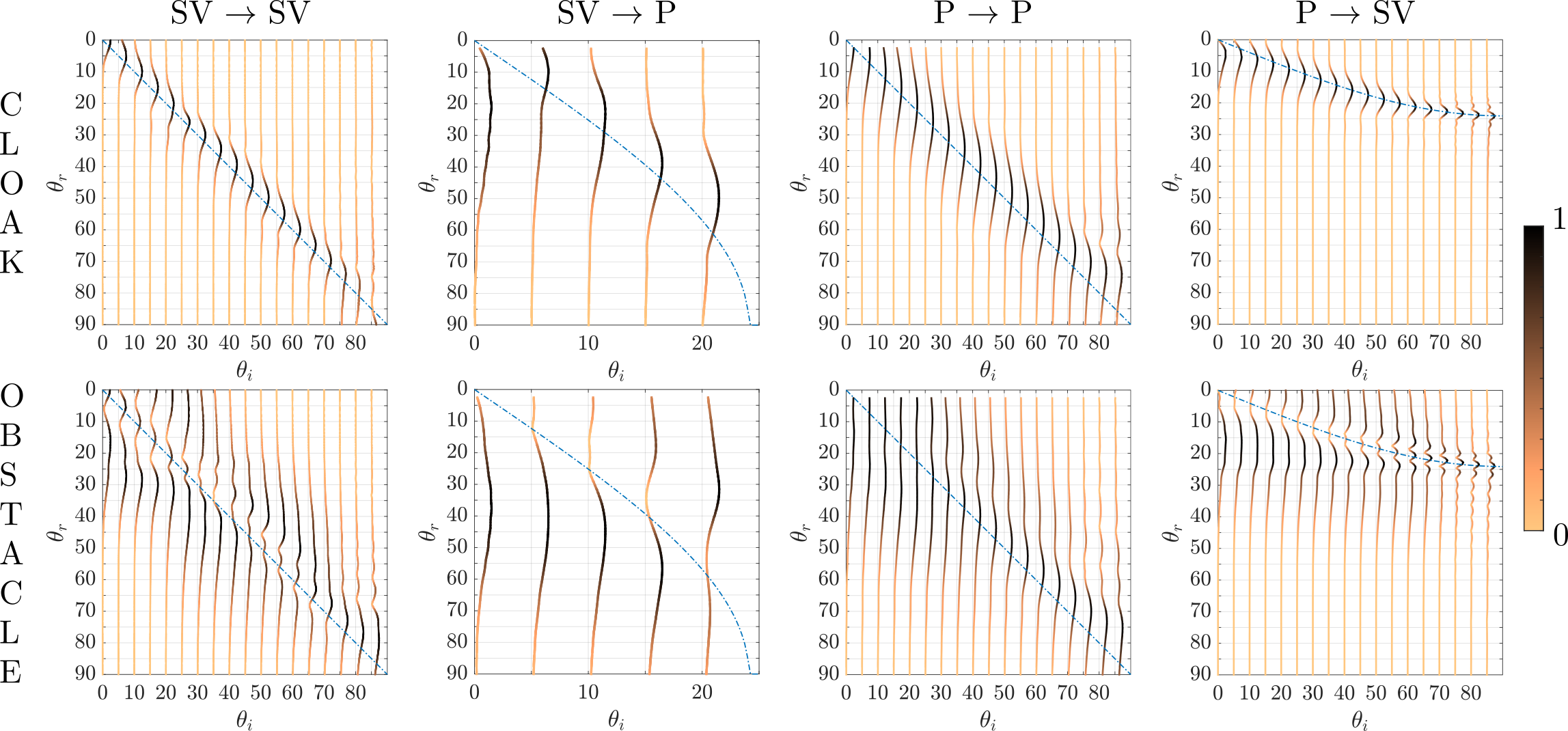}
     \caption{Comparison of the behavior of the cloak (\textit{above}) and  obstacle (\textit{below}) for different angles of incidence $\theta_i$, in terms of the FFT amplitude at $\kappa=\kappa_S$ and $\kappa=\kappa_P$, for shear and compression waves, respectively. Dashed lines represent expected angles of reflection according to Snell's law $c_r\sin\theta_r =c_i\sin\theta_i$. From left to right: 
     shear vertical wave reflected from incident shear vertical wave, P wave reflected from incident shear vertical wave (mode conversion), P wave reflected from incident P wave and shear vertical wave reflected from incident P wave (mode conversion).}
     \label{Figure3}
\end{figure*}
To validate the expected cloaking performances, we perform numerical finite element simulations of the full Navier equations based on the scattering formulation \citep{diatta2016scattering}. The mesh is designed to have at least twelve quadratic elements per minimum wavelength involved in the simulation to ensure convergence. The background material is  a polymer characterized by Young's modulus $E=3$ [GPa], Poisson's ratio $\nu=0.4$ and density $\rho=1190$ [kg/m$^3$]. 
Figure \ref{Figure2} visually displays the performance of the designed cloak at fixed frequency $f=200$ [kHz] for impinging SV and P waves at $\pi/4$ incidence. On the left side of Figure \ref{Figure2} 
the magnitude of the displacement field is depicted for the reference case, the plain obstacle, and the obstacle surrounded with the cloak, for a SV impinging wave. Each simulation is accompanied by the 2D fast Fourier transform (FFT) of the displacement field, that displays in wavenumber space the scattering directions. Each FFT is normalized with respect to its maximum, in order to clearly show all the directions of scattering. It is clearly seen that the cloak is able to conceal the presence of the Gaussian defect of the surface, which otherwise would scatter energy in almost every direction, with symmetric peaks at each side of the expected reference reflection direction. On the right side of  Figure \ref{Figure2}, the P wave case is shown. For both SV and P waves mode conversion can occur at a free surface, i.e, a transverse wave can be reflected as both a transverse and a longitudinal wave and, in the same way, a longitudinal wave can produce scattered waves with both polarizations. Such mode conversion obeys Snell's law, meaning that
\begin{equation}
    c_i\sin{\theta_i}=c_r\sin{\theta_r}
\end{equation}
with $\theta_i$, $\theta_r$ the incidence and reflection angles, and $c_i$, $c_r$ the speeds of the incident and reflected waves. As a consequence, the reflection angle associated to a SV wave will always be less than that associated to a P wave, $c_S$ being less than $c_P$. This in turn reflects that, for a 45 degree incident shear wave, no reflected P wave is expected, as seen from the reference and cloaked cases. The bare obstacle, instead, besides scattering transverse waves in several directions, also generates some mode conversion. Indeed the amplitude of the FFT is different from zero also for normalized $\kappa/\kappa_S\approx0.4$ which corresponds to the ratio of P to SV wavenumbers. The impinging P wave case clearly shows the conversion of polarizations, which can also be seen through the appearance of two strong spots in the FFT of the reference and cloaked cases. As in the other case, the obstacle scatters energy almost in all directions.\\
\begin{figure}[h]
 \centering
 \includegraphics[width=\textwidth,angle=0] {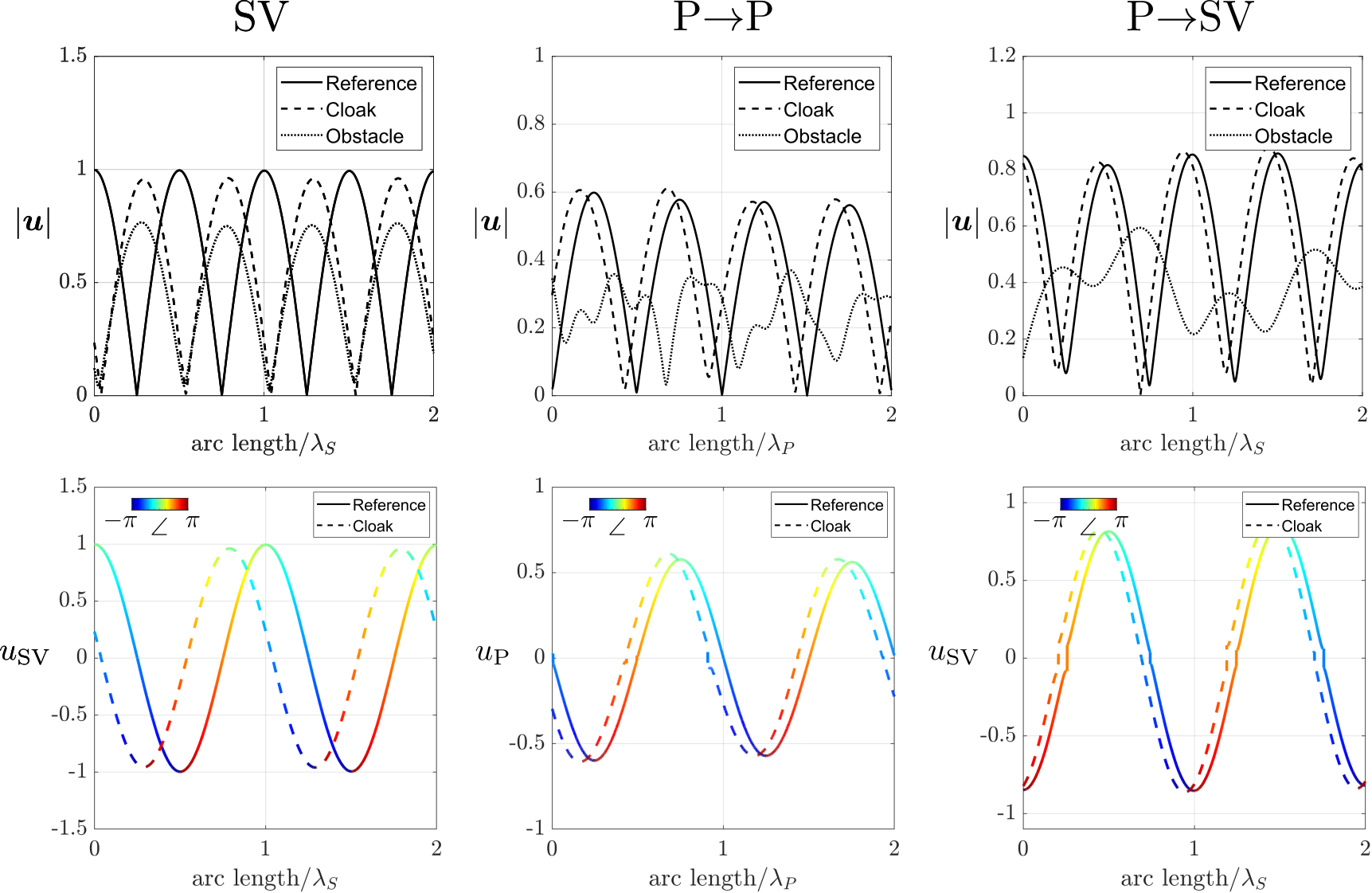}
 \caption{Comparison of the Reference, Obstacle and Cloak displacement fields as evaluated along the red lines shown in the bottom part of Figure \ref{Figure2}. In the top row a comparison is made in terms of the magnitude of the displacement vector, while in the bottom row the displacement is projected along the longitudinal (for P waves) and transversal (for SV waves) directions, for better comparison of amplitude and phase (depicted in colours). From left to right the SV wave reflected from incident SV wave is shown, then the P wave reflected from an incident P wave and finally the SV wave reflected from incident P wave.}
\label{Figure4}
\end{figure}
The dependence of the cloak performance on angle of incidence, $\theta_i$, is shown in figure \ref{Figure3}. For each $\theta_i$ the cloaked and bare obstacle are compared in terms of the distribution of the FFT calculated for wavenumbers corresponding to P and S waves (the white quarter of circles depicted in Figure \ref{Figure2}), which is plotted together with the related reflection direction $\theta_r=\arctan(\kappa_2/\kappa_1)$. This allows us  to assess whether  most of the signal is concentrated on the expected reflection angle (marked with dashed blue lines) or whether it is spread over a wide range of angles, and allows us to better appreciate the reduction of mode conversion. 

The dependence of cloaking performance with respect to frequency of operation and maximum width of the Gaussian defect 
 are completely defined by the indicators given by (\ref{magP}) and (\ref{magS}). Referring to the most demanding indicator, the calculated order of magnitude for the case shown in Figure \ref{Figure1} is:
\begin{equation}
    \frac{<\nabla \rho >/\rho_0}{\kappa_P}\approx 0.017 
\end{equation}
A thorough numerical sensitivity analysis of the performance of the cloak with respect to changes in this global indicator are given in the Supplementary Material.
\begin{figure}[h]
 \centering
 \includegraphics[width=\textwidth,angle=0] {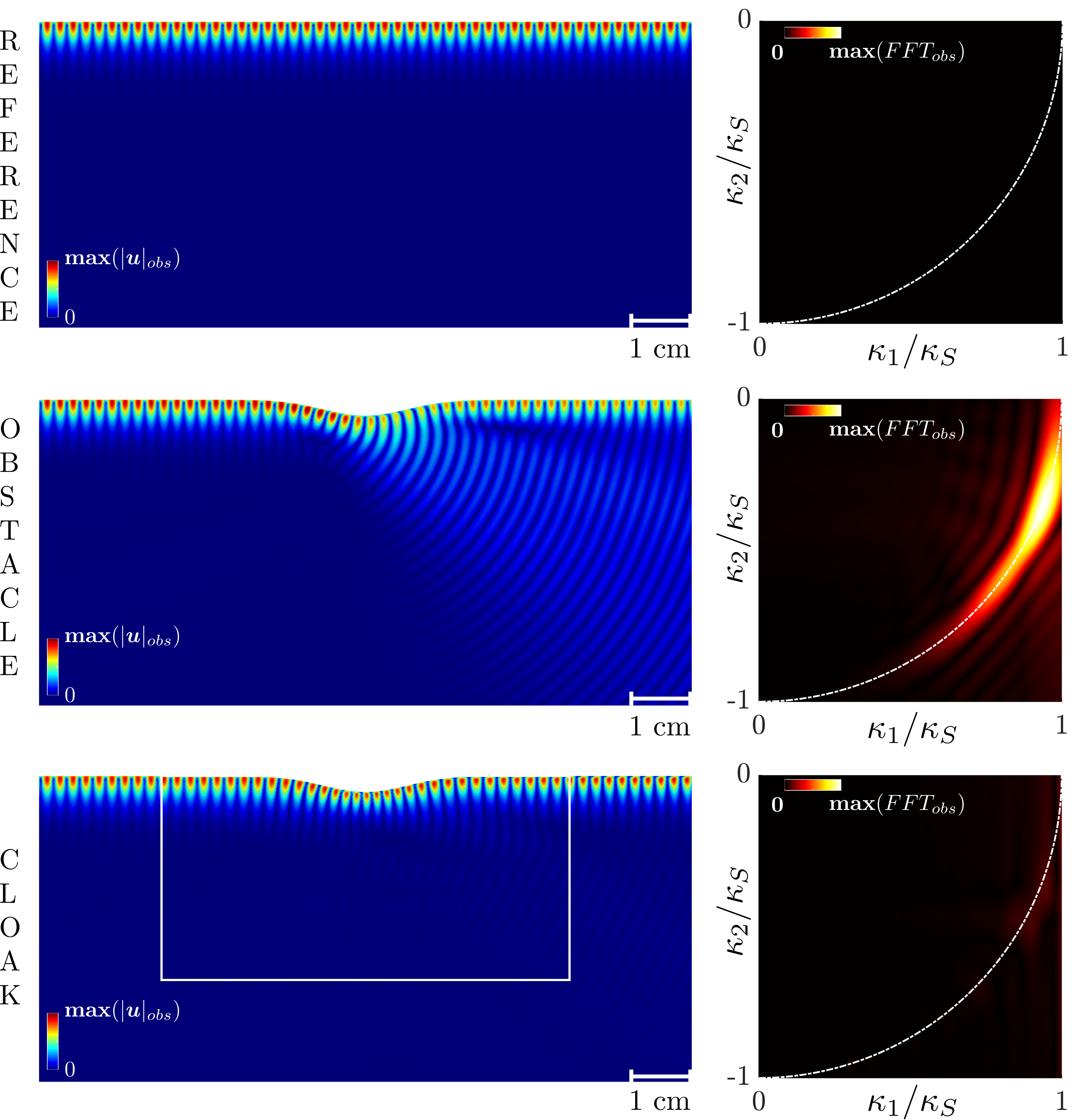}
 \caption{Mode conversion induced by surface obstacle on a Rayleigh wave. Scattering is reduced when the defect is surrounded by cloak.}
\label{Figure5}
\end{figure}
\\

As the cloak is finite, a phase shift is also expected between the reference and cloak solutions. This is hard to discern from the field plots shown in Figure \ref{Figure2} and hence we show solutions evaluated along the travelling direction of the scattered waves along the red lines shown in the lower panels of Figure \ref{Figure2} and plotted in Figure \ref{Figure4}. On the bottom line, from left to right we show a comparison between the cloak and reference solutions for an impinging SV wave, the reflected P wave from impinging P wave, and the reflected shear wave from impinging compressional one, respectively. The comparison is made in terms of the transversal displacement $u_S$ for the  shear waves and of longitudinal displacement $u_P$ for compressional waves. The color indicates the phase, highlighting the phase shift in the response of the cloaked obstacle. In the top part of the Figure, the same plots are shown in terms of magnitude of displacements, and the scattered field of the bare obstacle is also added for comparison.

\subsection{Carpet cloaking: Surface Rayleigh waves}
\label{sec:rayleigh}

Apart from incident purely compressional, P, or shear, S, waves, the cloak works well for other incident fields and we now turn to the important example of incident Rayleigh surface waves \cite{graff2012wave}. Figure \ref{Figure5} shows the comparison of the scattering behaviour of the bare obstacle and the cloaked obstacle when a Rayleigh wave tries to propagate across the deformed surface. In the uncloaked scenario substantial mode conversion occurs, and additionally part of the energy is scattered into the bulk of the material as a transverse SV wave, whereas the cloak considerably reduces such scattering, allowing the signal to propagate past the surface defect as if it was not there.
The critical measure of performance is not the angle of scattering, but is now the difference in total amount of energy that is converted into bulk waves and this can be visualised via the FFT and in Figure \ref{Figure5} the FFT of all the three cases have been normalized over the peak value of the obstacle case; this shows that remarkably little energy is scattered into the bulk by the cloaked solution. 

\section{Concluding remarks}
\label{sec:concluding}
We have demonstrated that near-cloaking is achievable with just density changes, and have quantified the error in doing so, and this opens the way to designing elastic cloaks for experiments and applications. Moreover, we have demonstrated the effectiveness of the approach that we advocate by showing that it is robust for all  angles of incidence and across a broad range of frequencies.


 Technically, using the decomposition of the Navier equation into a decoupled set of equations for scalar and vector potentials allows us to clearly isolate and quantify the extraneous terms introduced to marry the equation of the physical domain to the transformed equation.
This leads to a convenient adimensional parameter being introduced  to express the global dependence of performance with respect to working frequencies and local stretch imposed by the transformation. The best choice of material parameters that we found, to perform the near-cloak, comprises a gradient in the density only. This then naturally leads to viable materials for experiments as, for instance, metamaterials showing independent control of stiffness properties and density have been successfully manufactured \cite{kadic2014pentamode}, and can be employed to design inhomogeneous structures that are easily printed using gray-tone 3D direct laser writing.

\section*{Acknowledgement}
M.K. is grateful for support by the EIPHI Graduate School (contract ANR-17-EURE-0002) and by the French Investissements d'Avenir program, project ISITEBFC (contract ANR-15-IDEX-03).

F.B. acknowledges the Italian Ministry of Education, University and Research for the support provided through the Project "Department of Excellence LIS4.0 - Lightweight and Smart Structures for Industry 4.0”.


%
%
%

\section*{Supplementary Material}
To supplement the material in the main article we perform a parametric analysis of the performance of the near-cloaking. As explained in the main article, the cloak is achieved by assuming that additional terms connected to the slow variation of density, appearing in the equation of the physical domain are negligible. These terms depend both on the frequency of operation and on the gradient of density inside the cloak. Since the gradient of density is a local property of the cloak, we have introduced a global measure of performance with the following two non-dimensional indicators:
\begin{equation}
    \frac{<\nabla \rho >/\rho_0}{\kappa_P}=\frac{<1/n^2>c_P}{2\pi f}; \qquad \frac{<\nabla \rho >/\rho_0}{\kappa_S}=\frac{<1/n^2>c_S}{2\pi f} 
\end{equation}
which are obtained evaluating the ratio between the approximate magnitude of the extra terms that we are neglecting with respect to the other term in the equation. The smaller these ratios are, then the closer the performance of the cloak will be with respect to the reference case. From these equations, it can clearly be seen how the behavior of the cloak basically depends on two main factors: the shape of the defect, that prescribes the density distribution, and the frequency of the incident wave. 
\begin{figure}[h]
 \centering
 \includegraphics[width=\textwidth,angle=0] {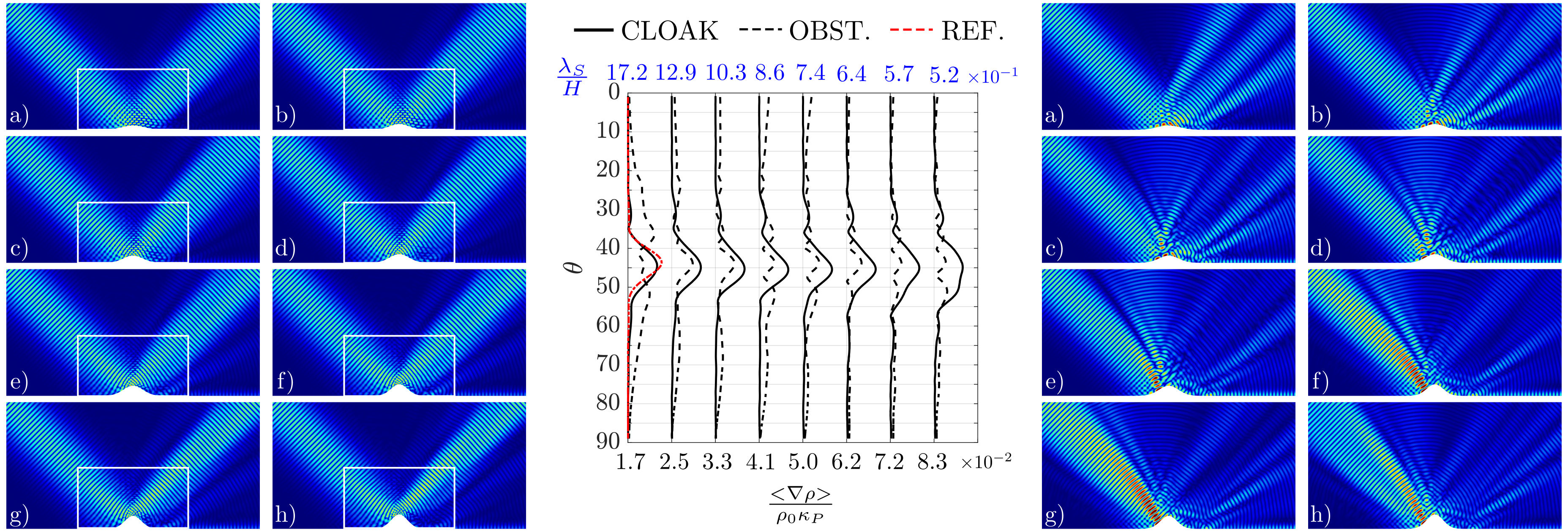}
 \caption{Effect of changing the size of the obstacle for impinging SV wave. The magnitude of the displacement field is shown for the cloak case (\textit{left}) and for the bare obstacle (\textit{right}). Letters a)-h) refer to increasing value of the adimensional parameter $<\nabla\rho>/\rho_0\kappa_P$. In the \textit{middle}, the evolution of the FFT of the scattered fields is shown as a function of the adimensional parameter governing the quality of the performance and the ratio wavelength/height of the bump. Solid black lines correspond to the cloak case, the dashed ones to the obstacle and the red one to the reference. Note that the reference case is the same for all the cases analyzed, thus is shown only one time.}
\label{Figure1}
\end{figure}

In Figure \ref{Figure1} a set of displacement fields for the cloak case (\textit{left}) and the obstacle case (\textit{right}) is depicted, in which the size of the defect is progressively increased. The corresponding increasing value of $<\nabla \rho >/(\rho_0\kappa_P)$ is shown in the black horizontal axis in the graph shown in the middle of Figure \ref{Figure1}, where the FFT of the scattered field along the circle corresponding to $\kappa=\kappa_S$ is compared between the reference, the obstacle and the cloak cases. For low values of the adimensional parameters, the cloak performs well in comparison with respect to the reference case, as seen by the peak of the FFT around $\theta_r=45^o$. As the size of the bump increases the performance of the cloak worsens as shown by the fact that the energy spreads. The response of the bare obstacle also shows increasing dispersion of energy in multiple scattering directions since the ratio between the wavelength and the height of the bump decreases (as shown with the blue horizontal axis): big objects are indeed more easily detected with shorter wavelengths.
\begin{figure}[h]
 \centering
 \includegraphics[width=\textwidth,angle=0] {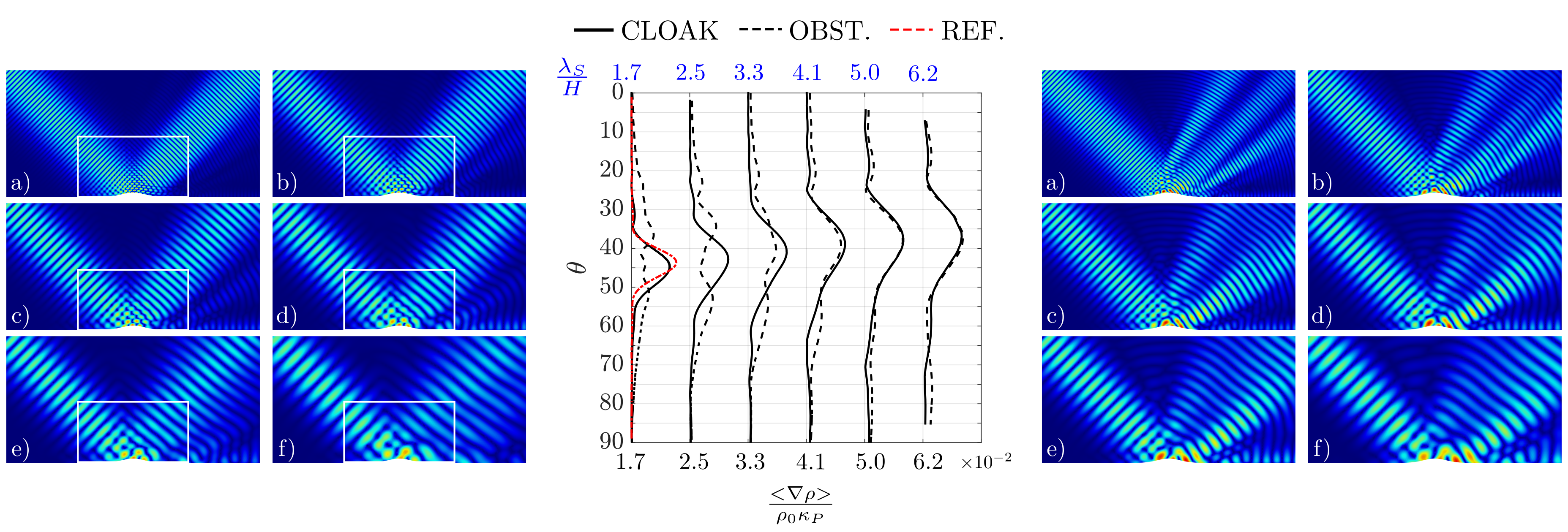}
 \caption{Effect of changing the working frequency for impinging SV wave. The magnitude of the displacement field is shown for the cloak case (\textit{left}) and for the bare obstacle (\textit{right}). Letters a)-f) refer to increasing value of the adimensional parameter $<\nabla\rho>/\rho_0\kappa_P$. In the \textit{middle}, the evolution of the FFT of the scattered fields is shown as a function both of the adimensional parameter governing the quality of performance and the ratio wavelength/height of the bump. Solid black lines correspond to the cloak case, the dashed ones to the obstacle and the red one to the reference.}
\label{Figure2}
\end{figure}
Figure \ref{Figure2} shows a similar analysis is conducted by keeping fixed the size of the bump, and decreasing the frequency of the impinging wave, in such a way to reproduce the same values of $<\nabla \rho >/(\rho_0\kappa_P)$ as the ones adopted in Figure \ref{Figure1}. As expected, things get worse for the cloak case as the wavelength increases. The blue horizontal axis shows that, in this case, the ratio between wavelength and size of the bump also increases, meaning that in the uncloaked case the performance would in principle get better by decreasing the frequency. For this reason, as the performance of the cloak decreases, the FFT of the cloak and bare obstacle converge to similar results.
\begin{figure}[h]
 \centering
 \includegraphics[width=\textwidth,angle=0] {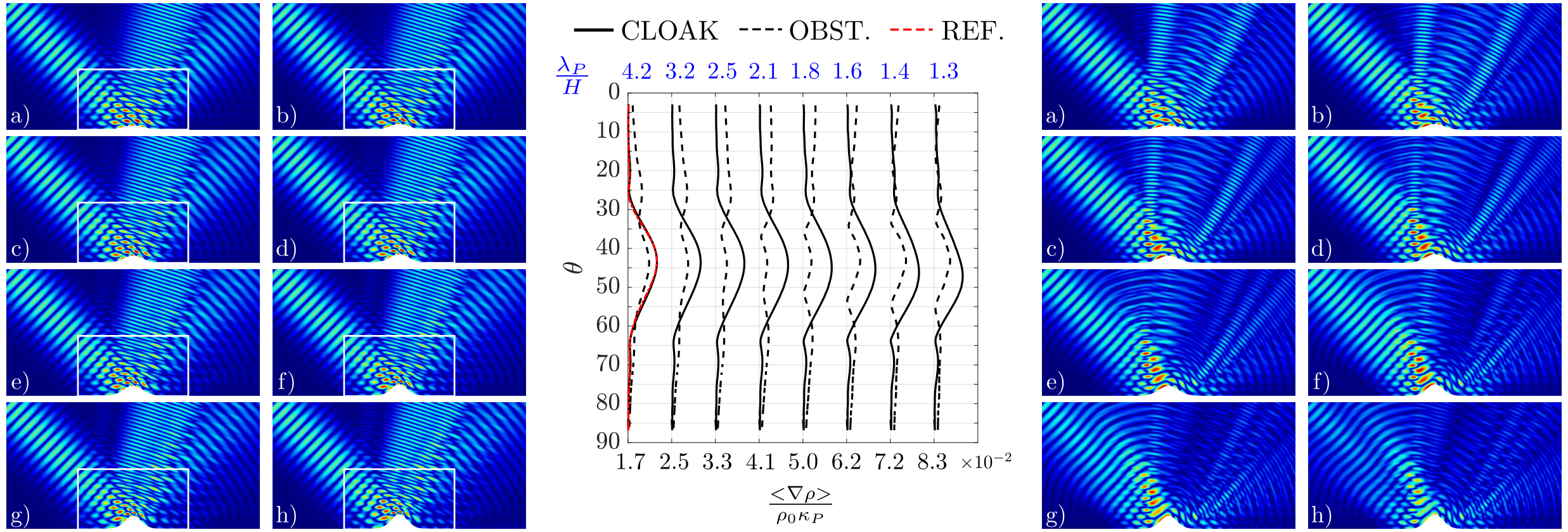}
 \caption{Effect of changing the size of the bump for a reflected P wave from an impinging P wave. The magnitude of the displacement field is shown for the cloak case (\textit{left}) and for the bare obstacle (\textit{right}). Letters a)-h) refer to increasing value of the adimensional parameter $<\nabla\rho>/\rho_0\kappa_P$. In the \textit{middle}, the evolution of the FFT of the scattered fields is shown as a function of the adimensional parameter governing the quality of the performance and the ratio wavelength/height of the bump. Solid black lines correspond to the cloak case, the dashed ones to the obstacle and the red one to the reference. Note that the reference case is the same for all the cases analyzed, thus is shown only one time.}
\label{Figure3}
\end{figure}

The same parametric study is performed for incident P waves also: in Figure \ref{Figure3} and \ref{Figure4} the FFT of the P waves scattered are shown alongside the displacement fields for variation of the size of the bump and of the working frequency, respectively. Figure \ref{Figure5} and \ref{Figure6} show the analysis for the SV wave scattered from the impinging P wave (mode conversion). The same trends underlined for the pure SV wave case can be observed also in these graphs.   
\begin{figure}[h]
 \centering
 \includegraphics[width=\textwidth,angle=0] {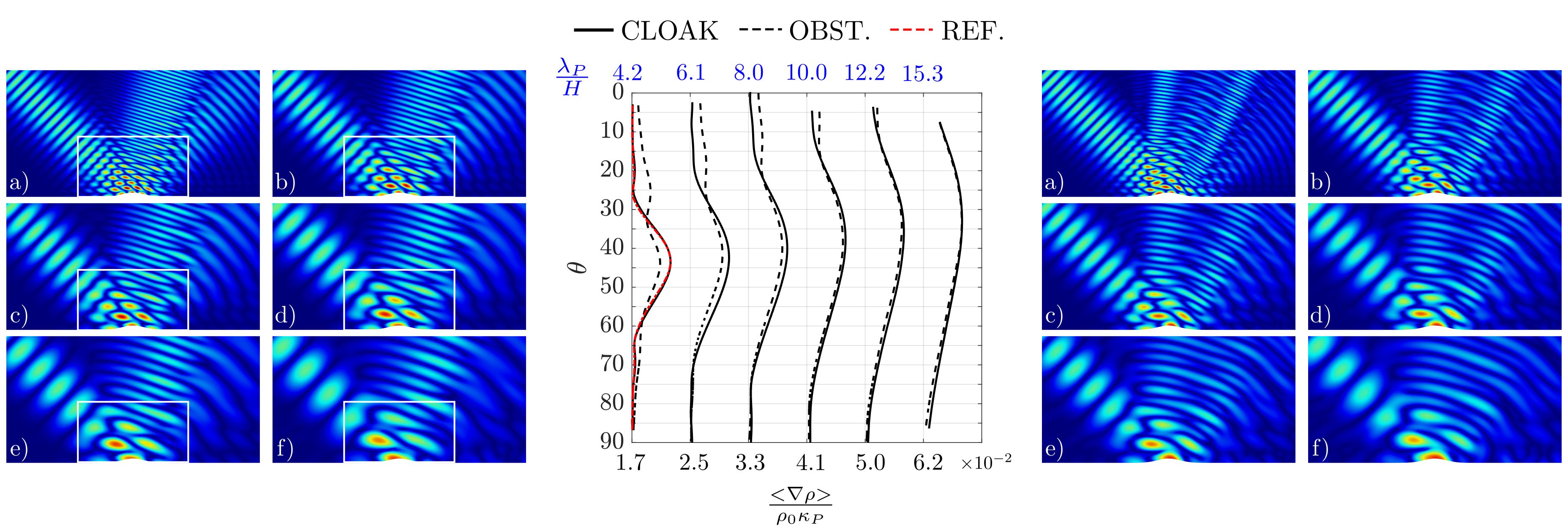}
 \caption{Effect of changing the working frequency for a reflected P wave from an impinging P wave. The magnitude of the displacement field is shown for the cloak case (\textit{left}) and for the bare obstacle (\textit{right}). Letters a)-f) refer to increasing value of the adimensional parameter $<\nabla\rho>/\rho_0\kappa_P$. In the \textit{middle}, the evolution of the FFT of the scattered fields is shown as a function both of the adimensional parameter governing the quality of performance and the ratio wavelength/height of the bump. Solid black lines correspond to the cloak case, the dashed ones to the obstacle and the red one to the reference.}
\label{Figure4}
\end{figure}
In the same way as for P and SV waves, Figure \ref{Figure7} and \ref{Figure8} deal with the Rayleigh waves.
\begin{figure}[h]
 \centering
 \includegraphics[width=\textwidth,angle=0] {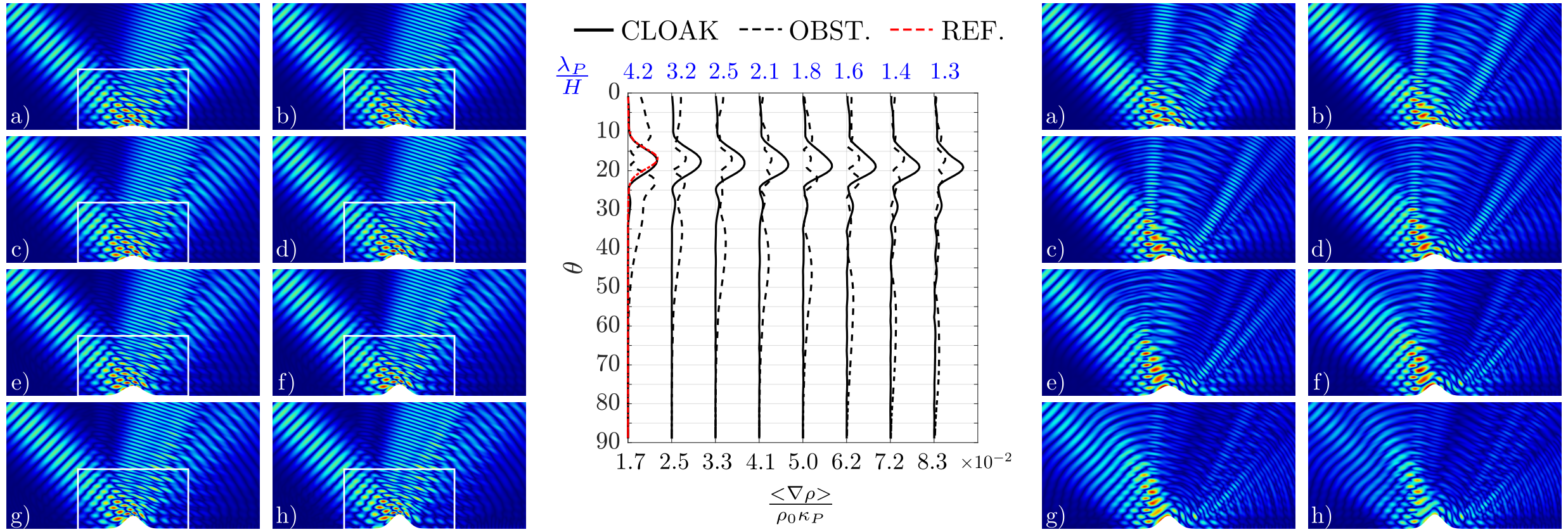}
 \caption{Effect of changing the size of the bump for a reflected SV wave from an impinging P wave. The magnitude of the displacement field is shown for the cloak case (\textit{left}) and for the bare obstacle (\textit{right}). Letters a)-h) refer to increasing value of the adimensional parameter $<\nabla\rho>/\rho_0\kappa_P$. In the \textit{middle}, the evolution of the FFT of the scattered fields is shown as a function both of the adimensional parameter governing the quality of performance and the ratio wavelength/height of the bump. Solid black lines correspond to the cloak case, the dashed ones to the obstacle and the red one to the reference.}
\label{Figure5}
\end{figure}
\begin{figure}[h]
 \centering
 \includegraphics[width=\textwidth,angle=0] {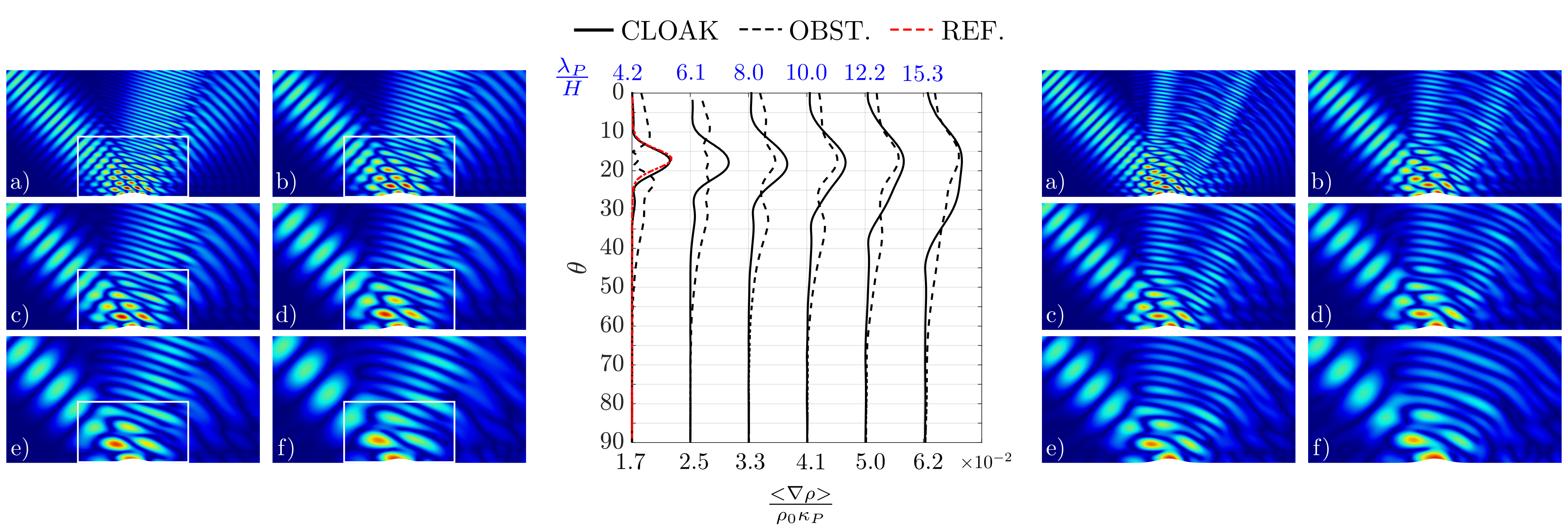}
 \caption{Effect of changing the working frequency for a reflected SV wave from an impinging P wave. The magnitude of the displacement field is shown for the cloak case (\textit{left}) and for the bare obstacle (\textit{right}). Letters a)-f) refer to increasing value of the adimensional parameter $<\nabla\rho>/\rho_0\kappa_P$. In the \textit{middle}, the evolution of the FFT of the scattered fields is shown as a function both of the adimensional parameter governing the quality of performance and the ratio wavelength/height of the bump. Solid black lines correspond to the cloak case, the dashed ones to the obstacle and the red one to the reference.}
\label{Figure6}
\end{figure}
\begin{figure}[h]
 \centering
 \includegraphics[width=\textwidth,angle=0] {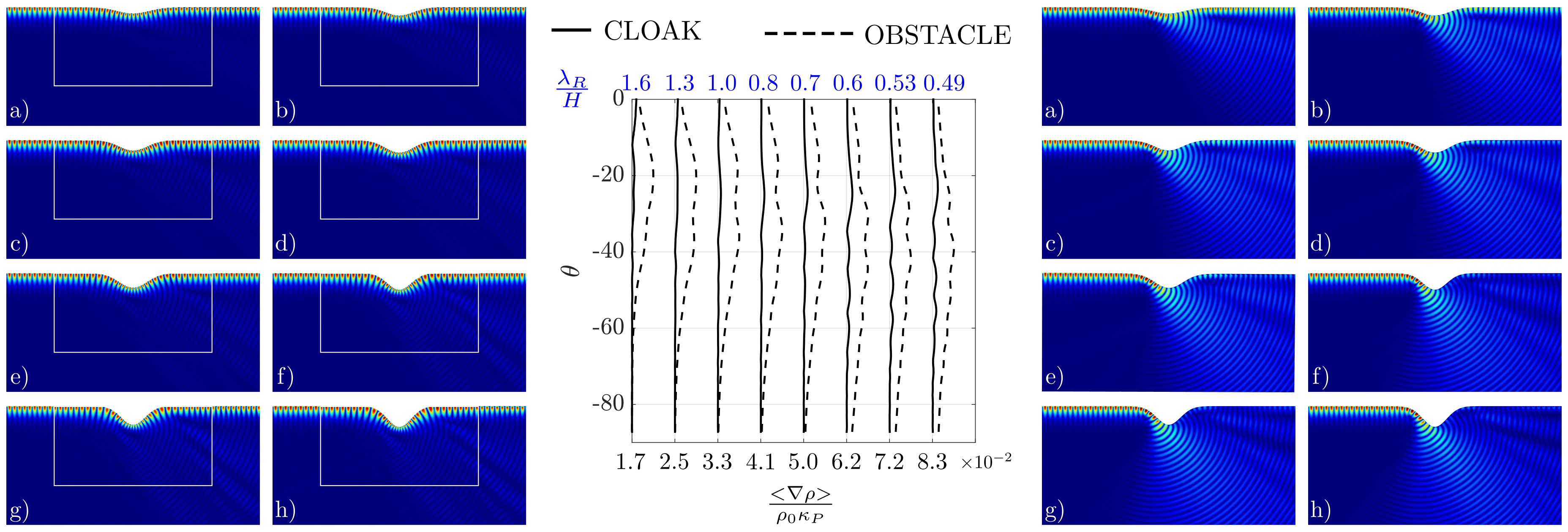}
 \caption{Effect of changing the size of the bump for Rayleigh waves. The magnitude of the displacement field is shown for the cloak case (\textit{left}) and for the bare obstacle (\textit{right}). Letters a)-h) refer to increasing value of the adimensional parameter $<\nabla\rho>/\rho_0\kappa_P$. In the \textit{middle}, the evolution of the FFT of the scattered fields is shown as a function both of the adimensional parameter governing the quality of performance and the ratio wavelength/height of the bump. Solid black lines correspond to the cloak case while the dashed ones to the obstacle.}
\label{Figure7}
\end{figure}
\begin{figure}[h]
 \centering
 \includegraphics[width=\textwidth,angle=0] {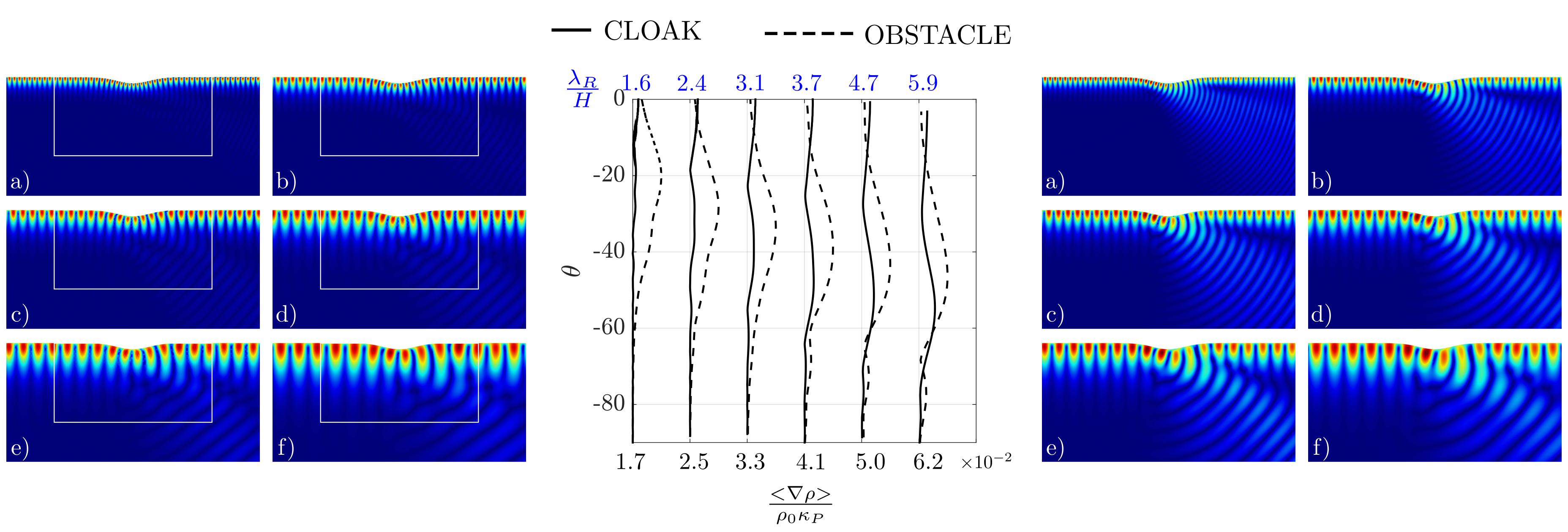}
 \caption{Effect of changing the working frequency for Rayleigh waves. The magnitude of the displacement field is shown for the cloak case (\textit{left}) and for the bare obstacle (\textit{right}). Letters a)-f) refer to increasing value of the adimensional parameter $<\nabla\rho>/\rho_0\kappa_P$. In the \textit{middle}, the evolution of the FFT of the scattered fields is shown as a function both of the adimensional parameter governing the quality of performance and the ratio wavelength/height of the bump. Solid black lines correspond to the cloak case while the dashed ones to the obstacle.}
\label{Figure8}
\end{figure}

\end{document}